\documentclass{appolb}
\usepackage{cite,here,epsfig}

\begin{document}
\title{ Proton-proton collisions at production thresholds
}
\author{
         P.~Moskal$^{1,2}$,
         H.-H.~Adam$^3$,
         A.~Budzanowski$^4$,
         D.~Grzonka$^2$,
         L.~Jarczyk$^1$,
         A.~Khoukaz$^3$,
         K.~Kilian$^2$,
         M.~K\"{o}hler$^5$,
         P.~Kowina$^6$,
         N.~Lang$^3$,
         T.~Lister$^3$,
         W.~Oelert$^2$,
         C.~Quentmeier$^3$,
         R.~Santo$^3$,
         G.~Schepers$^{2,3}$,
         T.~Sefzick$^2$,
         S.~Sewerin$^2$\footnote{present address: The Svedberg Laboratory S-75121 Uppsala, Sweden},
         M.~Siemaszko$^6$,
         J.~Smyrski$^1$,
         M.~Soko{\l}owski$^1$,
         A.~Strza{\l}kowski$^1$,
         M.~Wolke$^2$,
         P.~W{\"u}stner$^5$,
         W.~Zipper$^6$
 \address{ $^1$ Institute of Physics, Jagellonian University, PL-30-059 Cracow, Poland        \\
           $^2$ IKP, Forschungszentrum J\"{u}lich, D-52425 J\"{u}lich, Germany                \\
           $^3$ IKP, Westf\"{a}lische Wilhelms--Universit\"{a}t, D-48149 M\"{u}nster, Germany \\
           $^4$ Institute of Nuclear Physics, PL-31-342 Cracow, Poland                        \\
           $^5$ ZEL,  Forschungszentrum J\"{u}lich, D-52425 J\"{u}lich,  Germany              \\
           $^6$ Institute of Physics, University of Silesia, PL-40-007 Katowice, Poland       \\
         }
}
\maketitle
\begin{abstract}
   Recent results obtained by the COSY~--~11 collaboration 
   concerning the production of $\eta$ and $\eta^{\prime}$ mesons in the 
   $pp \rightarrow pp Meson$ reaction   are presented.
   A comparison of the 
   production amplitude for the $\pi^{0}$, $\eta$ and $\eta^{\prime}$ mesons
   at the same phase space volume allows to conclude that the  
   proton~-~$\eta^{\prime}$ interaction is in the order of,
   or smaller than, the proton-$\pi^{0}$ one.  \\
   A total cross section determined
   in a preliminary analysis of the data of  elementary kaon and antikaon production
   via the $pp \rightarrow pp K^{+}K^{-}$ reaction  measured at excess energy of $Q$~=~17~MeV
   is reported.
\end{abstract}
  
\section{Introduction}
  In the last decade 
a copious set of data on the close-to-threshold production of mesons 
$\pi^{0}$,
$\eta$,
 and $\eta^{\prime}$ in the collisions of protons
has been collected
at the high precision accelerators in Bloomington, Uppsala, Saclay and J\"ulich.
The quality of the determined energy dependence of the total cross sections
for the $pp\rightarrow pp\pi^{0}$~\cite{bondar95,meyer92,meyer90}, 
$pp\rightarrow pp\eta$~\cite{smyrskipl,calenpdeta,caleneta,chiavassa,bergdolt}, 
and $pp\rightarrow pp\eta^{\prime}$~\cite{moskalpl,hiboupl,moskalprl}
reactions enables investigations of the microscopic description 
of the primary production 
mechanism~\cite{wilk,bernard,hernandez,hanhartpl,shyammosel,horowitz,pena,batinic,vetter,kleefeld,nakayama,bass2,gedalin,sibirtsev} 
and the interaction of protons with the created meson~\cite{pena,baruej,moskalpl2}. 
Here a 
special emphasis is given to the still unknown
interaction of protons with the $\eta^{\prime}$ meson, 
which can not be studied directly in the elastic $\eta^{\prime}$-proton scattering,
due to the short life time of this meson. 
This issue will be discussed in the next section  where the qualitative phenomenological
analysis will be presented which results in the rough estimation of an upper limit for the 
proton-$\eta^{\prime}$ scattering length.
In the third section preliminary results concerning the studies of the open strangeness production
via the $pp\rightarrow ppK^{+}K^{-}$ 
reaction close to the production threshold will be overviewed.   

 Since the subject of this report covers
only a part of the COSY~--~11 activity  the interested reader is encouraged for further reading 
of an unexpected large difference, observed recently, 
in the close-to-threshold $K^{+}$ meson production depending whether it is associated 
with a~$\Sigma^{0}$ or a~$\Lambda$ hyperon~\cite{sewerinprl,balewskipl,sewerinnp}.

\section{S-wave proton-$\eta^{\prime}$ interaction}

\begin{figure}[H]
 \unitlength 1.0cm
  \begin{picture}(12.2,12.0)
    \put(2.0,0.0){
           \epsfig{figure=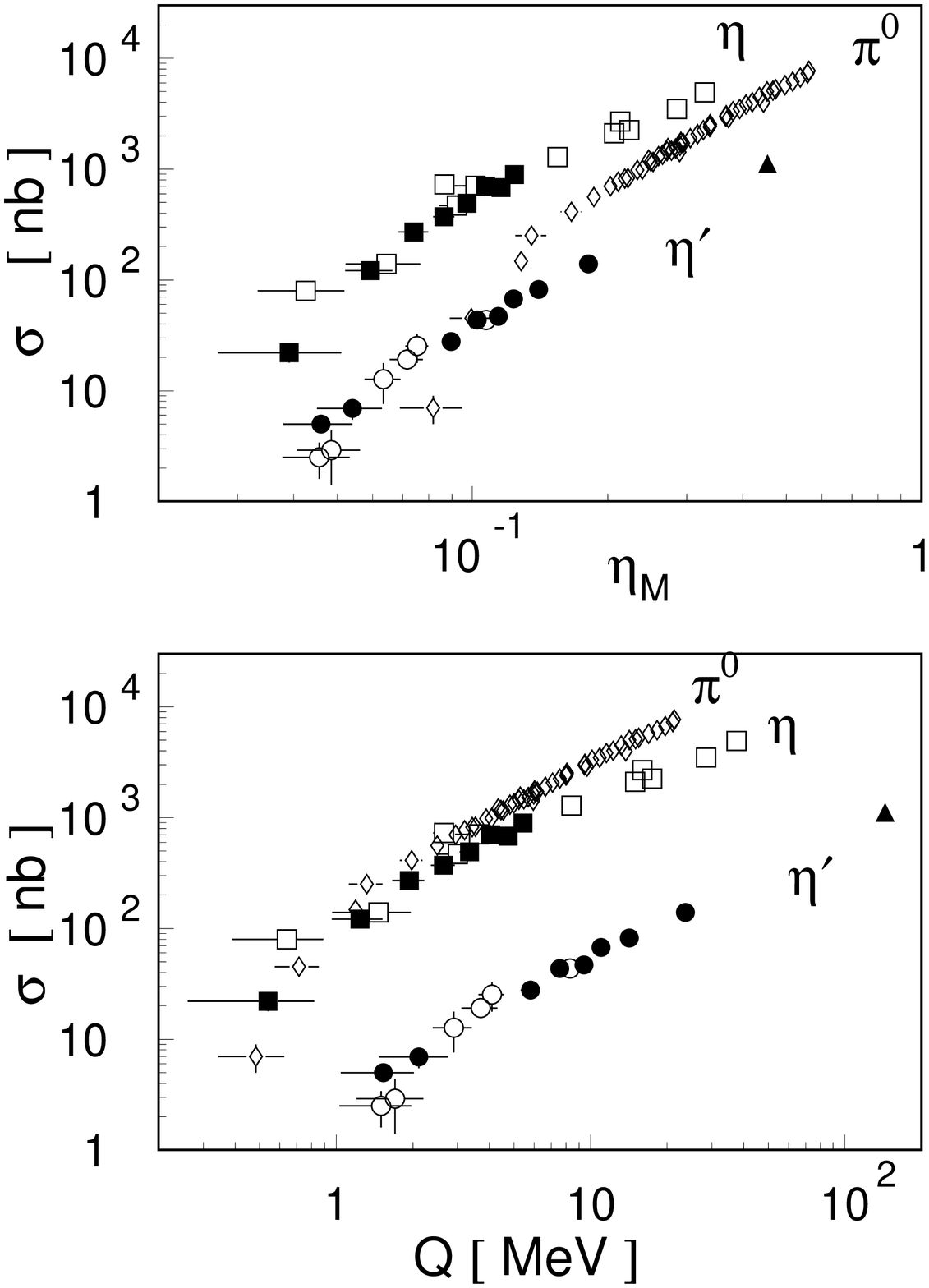,height=13.0cm,width=9.0cm,angle=0}
    }
    \put(10.7,6.6){
          { a)}
    }
    \put(10.7,1.0){
          { b)}
    }
  \end{picture}
  \caption{
           Total  cross sections for the reactions
           $pp \rightarrow pp \pi^{0}$ (diamonds~\protect\cite{bondar95,meyer92,meyer90}), \ \
           $pp \rightarrow pp \eta$ (squares~\protect\cite{smyrskipl,caleneta,chiavassa,bergdolt,hiboupl}), and \ \
           $pp \rightarrow pp \eta^{\prime}$ (circles~\protect\cite{moskalpl,hiboupl,moskalprl}, 
                                              triangle~\protect\cite{salabura})  \ \ \ \
            a) versus the maximum center-of-mass meson momentum normalized to the meson mass $\eta_{M}$, and  \ \ \ \
             b) as a function of the excess energy $Q$. \ \ \
           The filled squares and circles indicate  recent COSY~--~11 results~\protect\cite{smyrskipl,moskalpl},
           and the filled triangle was reported on this conference~\protect\cite{salabura}.
  }
  \label{cross_q_eta}
\end{figure}

 Trying to compare the total cross section for the close-to-threshold production of different mesons 
one has to find an appropriate kinematical variable. 
Usually, the total cross section is presented as a function of the dimensionless
parameter~$\eta_{M}$~\cite{bondar95,meyer92,machner}~\footnote{
      In order to avoid ambiguities with the abbreviation for the eta-meson, we introduce an additional
      suffix M  for this parameter, which usually is called  $\eta$ only.
      },
which is
defined as the maximum center-of-mass meson momentum in  units of meson
mass~\begin{math}\begin{displaystyle}(\eta_{M} = \frac{q_{max}}{M})\end{displaystyle}\end{math},
or as a function of the excess energy~$Q$~\cite{caleneta,bergdolt,hiboupl}.
 In Figure~\ref{cross_q_eta}a  the total cross sections for the
reactions $pp\rightarrow pp\pi^{0}$,  $pp\rightarrow pp\eta$, and  $pp\rightarrow pp\eta^{\prime}$
are compared versus the parameter $\eta_{M}$ and in Figure~\ref{cross_q_eta}b versus the excess energy.
One immediately notices the qualitative difference between both representations. 
For example, the $\eta$ meson production cross section exceeds the $\pi^{0}$ cross section
by  a factor of 2 and more using $\eta_{M}$,
whereas the $\pi^{0}$ meson cross section is always larger than the $\eta$ one  when  the $Q$ scale is employed.
 To find a proper variable for the comparison of the cross sections for mesons of significantly
different masses we recall a definition of the total cross section, which is just the 
integral over phase space 
of the squared  transition matrix element normalized to the
incoming flux factor~F:
 \begin{equation}
  \sigma_{pp\rightarrow ppX}=
  \frac{ 1 }{ F } \displaystyle \int dV_{ps} \ |M_{pp\rightarrow ppX}|^{2},
  \label{eqcrossform}
\end{equation}
where
 $X$ stands for the $\pi^{0},\eta$ or $\eta^{\prime}$ meson,
 $V_{ps}$ denotes the phase space volume, and
$F$~=~\begin{math}2~\ (2\pi)^{5}~\sqrt{s~\ (s~-~4~\ ~m_{p}^{2})}\end{math}~\cite{byckling},
with $s$ being the square of the total energy in the center-of-mass frame.

This definition 
suggests that a natural variable for comparing  the total cross sections for different
mesons may be the
volume of available phase space~\cite{moskalpl2}. 
Note that in case of the same  dynamics (transition matrix element) for the production 
of two different mesons  we would obtain identical values for the total cross section
as a function of $V_{ps}$ independently from  the produced meson masses, which would not be the case
when the variables $\eta_{M}$ or $Q$ would have been employed.

 Now, in order to study the proton-$\eta^{\prime}$ interaction we will employ two assumptions~\cite{moskalpl2},
which were lively discussed during this conference~\cite{kleefeld2,niskanenpl,baru2,hannak}:
  
  1) \ in analogy with the {\em Watson-Migdal} approximation~\cite{wats52} for two body processes,
we will assume that the complete transition amplitude for a production process
$M_{pp\rightarrow ppX}$ factorizes approximately as:
\begin{center}
  $ |M_{pp \rightarrow ppX}|^{2}   \approx  |M_{0}|^{2} \cdot |M_{FSI}|^{2} \cdot ISI, $
\end{center}
where  $M_{0}$ represents the total
production amplitude, $M_{FSI}$  describes the elastic interaction among particles
in the exit channel, and $ISI$ denotes the reduction factor
due to the interaction of the colliding protons.
    This factorization, however, is valid 
    only as long as the energy dependence of the total cross section
    is considered~\cite{niskanenpl,baru2,hannak}.
 
 2) \  we will assume also that in the exit channel only the proton-proton interaction
     is significant ($|M_{FSI}|^{2}$ = $|M_{pp \rightarrow pp}|^{2}$).
\begin{figure}[H]
 \unitlength 1.0cm
  \begin{picture}(12.2,10.5)
    \put(3.0,0.0){
       \epsfig{figure=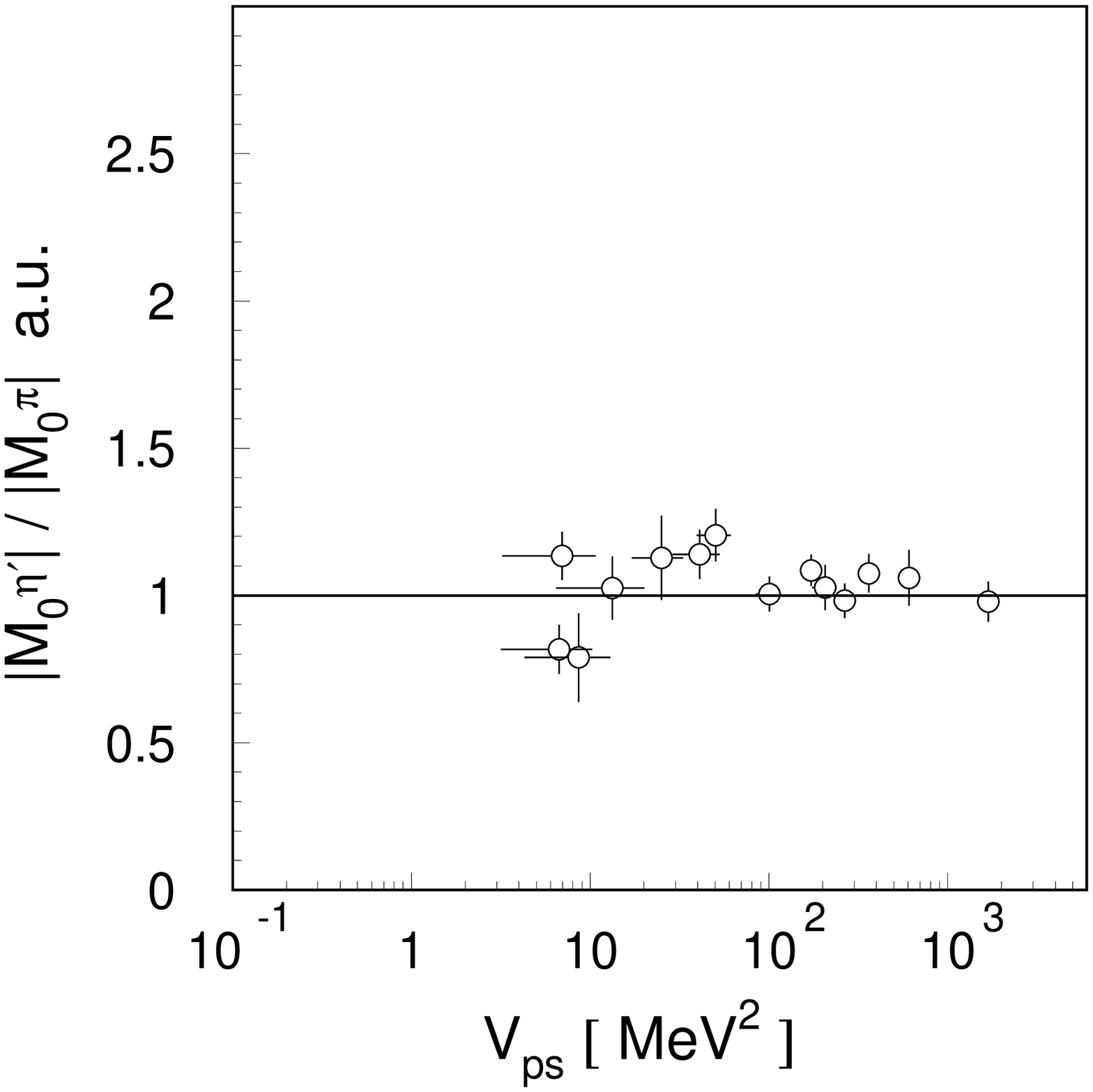,height=5.8cm,width=7.0cm,angle=0}
    }
       \put(9.5,1.5){
          { b)}
       }
    \put(3.0,4.5){
       \epsfig{figure=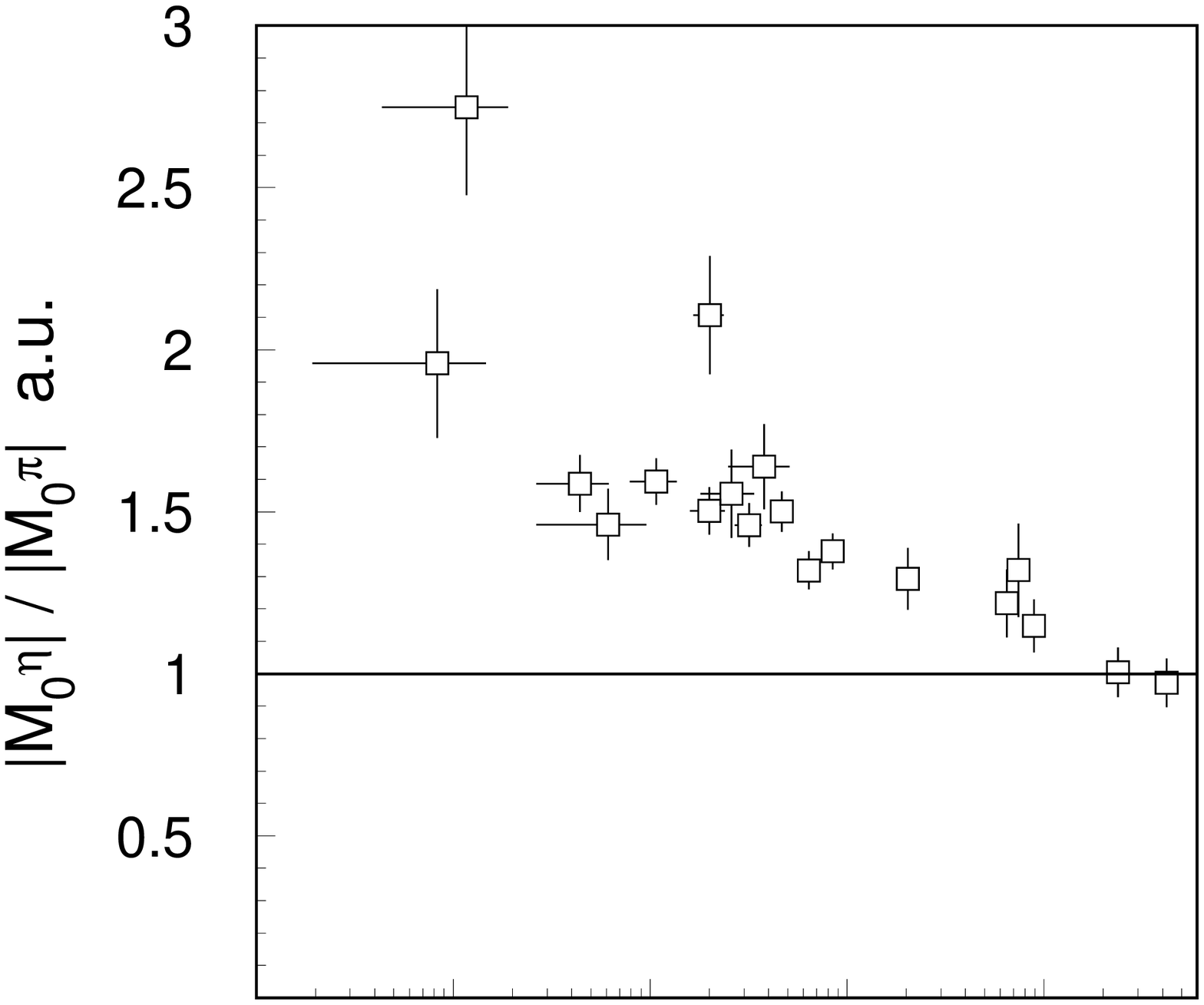,height=5.8cm,width=7.0cm,angle=0}
    }
       \put(9.5,6.0){
          { a)}
       }
  \end{picture}
  \caption{ \small
            The ratios of \ \ a) $|M^{\eta}_{0}|/|M^{\pi^{0}}_{0}|$
             and \ \ b) $|M^{\eta^{\prime}}_{0}|/|M^{\pi^{0}}_{0}|$
             extracted from the data, assuming the
             $pp-FSI$ enhancement factor of references~\cite{shyammosel,watson}.
        }
\label{mmnisk}
\end{figure}

These two assumptions enable us to derive from the measured cross sections 
the  phase space dependence of $|M^{\pi^{0}}_{0}|^{2}$, $|M^{\eta}_{0}|^{2}$, and
$|M^{\eta^{\prime}}_{0}|^{2}$~\cite{moskalpl2,moskalaip},
since the $M_{pp \rightarrow pp}$ amplitude is known and the $ISI$ factor can be calculated
according to the formula from reference~\cite{hannak}. However, there is no unequivocal
description for the  $|M_{pp \rightarrow pp}|^{2}$
enhancement factor~\cite{moskalpl2}. 
Therefore, to minimize ambiguities resulting
from  this uncertainties,
we consider
the ratios $|M^{\eta}_{0}|/|M^{\pi^{0}}_{0}|$ and $|M^{\eta^{\prime}}_{0}|/|M^{\pi^{0}}_{0}|$~\cite{moskalpl2},
which normalize the transition amplitude for $\eta$ and $\eta^{\prime}$ 
 to the one for $\pi^{0}$ production $|M^{\pi^{0}}_{0}|$.
This should be independent of the model used for the determination of $|M_{pp\rightarrow pp}|^{2}$,
and will allow an estimate of
the relative strength of the $\pi^{0}$-proton and $\eta(\eta^{\prime})$-proton interactions.
Indeed, we examined that within the errors the ratio $|M^{\eta(\eta^{\prime})}_{0}|/|M^{\pi^{0}}_{0}|$
does not depend on the model used for  $|M_{pp\rightarrow pp}|^{2}$.
As an example, in Figure~\ref{mmnisk} we show this ratio as obtained
from the  amplitude $|M_{pp\rightarrow pp}|^{2}$ taken from references~\cite{shyammosel,watson}.
Figure~\ref{mmnisk}a shows an increasing strength
of $|M_{0}|$  for the $\eta$ production at low $V_{ps}$,
indicating a strong $\eta$-proton FSI, as was discussed previously for the cross section ratio
by Cal\'{e}n et al.~\cite{caleneta}.
Note also that  the ratio for the $\eta^{\prime}$ meson
is constant over the phase space range considered (Figure~\ref{mmnisk}b).
This observation, and the fact that theoretical
calculations predict  the primary production amplitude to be constant
within a few per cent~\cite{nakayama,gedalin}
independently  from the mechanism assumed, allows us to conclude that
the $\eta^{\prime}$-proton scattering parameters are 
in the order of,
   or smaller than, the proton-$\pi^{0}$ ones.

\section{Kaon and antikaon production}

 Two years ago at the MESON'98 we presented  upper limits for the total cross section
of the $pp\rightarrow pp K^{+}K^{-}$ reaction~\cite{moskalacta}. 
 At present due to the gained statistics
and the understanding of the background we are pleased to present an absolute value for the
total cross section at an excess energy of $Q$~=~17~MeV.
The primordial motivation
for studying this reaction was presented already ten years ago at this conference  hall  by W. Oelert~\cite{oelert}.
It concerns the study of the  $K^{+}K^{-}$ interaction and the investigation of the structure of the $f_{0}$(980) meson, 
which is still discussed to be either the usual $q\bar{q}$, the exotic state $qq\bar{q}\bar{q}$ 
or a strongly bound $K\bar{K}$ molecule.

\begin{figure}[H]
 \unitlength 1.0cm
  \begin{picture}(12.0,7.0)
    \put(2.0,0.0){
           \epsfig{figure=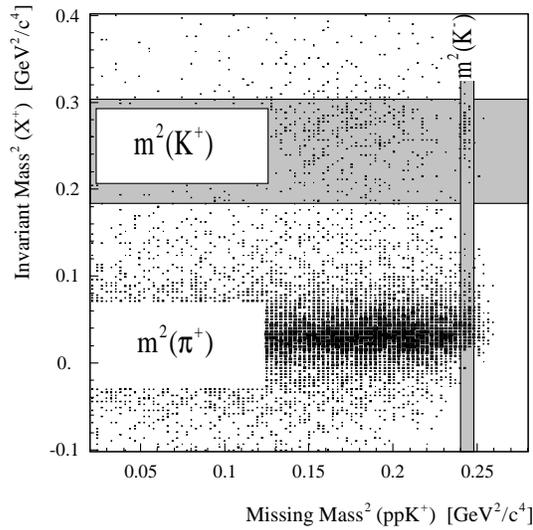,height=7.0cm,angle=0}
    }
  \end{picture}
  \caption{
     Results of the preliminary analysis~\cite{quentmeier}:
     Invariant mass of one out of three positively charged particles which was not identified as a proton
     versus the missing mass of an assumed ($ppK^{+}$)-subsystem. 
     The shaded areas, centered around the mass of the kaon, indicate  three
     standard deviations of the experimental resolution.
  }
  \label{inv_vs_mm_3356}
\end{figure}

\begin{figure}[H]
 \unitlength 1.0cm
  \begin{picture}(12.0,6.5)
    \put(2.5,0.0){
           \epsfig{figure=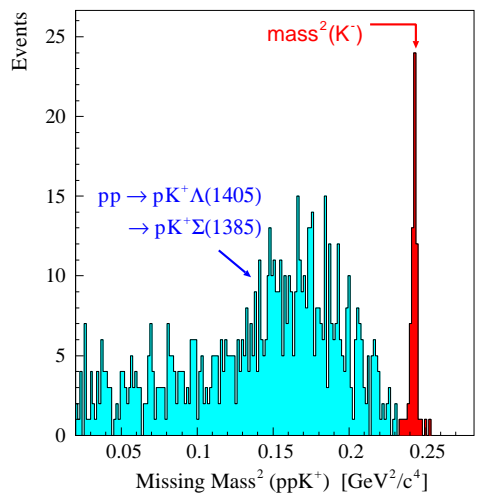,height=6.5cm,angle=0}
    }
  \end{picture}
  \caption{
      Result of a preliminary analysis~\cite{quentmeier}:
      Missing mass with respect to an identified  ($ppK^{+}$)-subsystem as measured
      at excess energy of $Q$~=~17~MeV.
  }
  \label{mmppk}
\end{figure}

 In order to identify this reaction 
 the four-momentum vectors for three positively charged particles are determined~\cite{wolkeaip}.
Figure~\ref{inv_vs_mm_3356} shows the preselected data where two of the positively charged particles
are identified as protons. On the vertical axis the measured mass of the third registered particle
is plotted as a function of the mass of an unobserved system. In the case of the $pp \rightarrow ppK^{+}K^{-}$
reaction both the invariant mass of the third particle and the missing mass~--~with respect to the identified
($ppK^{+}$) subsystem~--~should 
correspond to the mass of the kaon. 
Already at this level of analysis
the group of events corresponding to the $pp \rightarrow ppK^{+}K^{-}$ reaction can be recognized.
 The projection of events contained in the horizontal shaded area onto the missing mass axis 
reveals a clear  signal originating from the  $K^{+}K^{-}$ meson pair production, as presented in Figure~\ref{mmppk}.
The much broader structure seen on the left side of the peak is due to the $K^{+}$ meson production 
associated with the hyperon resonances $\Lambda$(1405) or $\Sigma$(1385) \ \ 
(eg. $pp\rightarrow pK^{+}\Lambda(1405)\rightarrow pK^{+}\Sigma\pi 
\rightarrow pK^{+}\Lambda\gamma\pi \rightarrow pK^{+}p\pi\gamma\pi$). 
In this case the missing mass of the pp$K^{+}$
system corresponds to the invariant mass of the ($\pi\pi\gamma$) subsystem.
  Demanding an additional signal in a silicon pad detector~\cite{brauksiepenim} at the position expected for the
$K^{-}$ meson the background is reduced by more than one order of magnitude~\cite{quentmeier}. 
This  additional requirement diminishes
the signal from
the $pp \rightarrow ppK^{+}K^{-}$ by about 50~$\%$ only~\cite{quentmeier}, which is understood by the decay 
of the $K^{-}$ meson on its way to the dedicated silicon detector.

The preliminary analysis of the data taken at the excess energy of $Q$~=~17~MeV 
results in a cross section value of 2.1~$\pm$~0.8~nb~\cite{quentmeier,wolkeaip,listerikp,khoukaznp}.  
When compared to the value 200~$\pm$~11~$\pm$~80~nb determined at $Q$~=~111~MeV~\cite{balestrapl},
one observes that the cross section for the  production of the $K^{-}$ 
meson in the elementary proton-proton collisions  increases much stronger than the corresponding one
for the $K^{+}$~\cite{balewskipl,bilgerpl} meson.

\end{document}